\documentclass[12pt]{iopart}

\usepackage{iopams}  
\usepackage{graphicx}
\begin{document}

\title[Frame representations of qudit quantum mechanics]{Frame representations of qudit quantum mechanics}

\author{Nicolae Cotfas}

\address{Faculty of Physics, University of Bucharest,\\
P.O. Box MG-11, 077125 Bucharest, Romania.}
\ead{nicolae.cotfas@unibuc.ro, https://unibuc.ro/user/nicolae.cotfas/}
\vspace{10pt}
\begin{indented}
\item[]\today
\end{indented}

\begin{abstract}
There exist many attempts to define a Wigner function for qudits, each of them coming with its advantages and limitations. 
The existing finite versions have simple definitions, but they are artificial in their construction and do not allow an intuitive state analysis. 
The continuous versions have more complicated definitions, but they are similar to the original Wigner function and allow a visualization of the quantum states. 
The version based on the concept of tight frame we present is finite, but it has certain properties and applications similar to those of continuous versions. 
Based on the frame representation, we present several graphical representations of qubit states, and define two new parameters concerning them. We show that, from a mathematical point of view, the qubit is the orthogonal projection of qutrit.
\end{abstract}

%\maketitle

\section{Introduction}

In the case of a qudit described by a complex $d$-dimensional Hilbert space $\mathcal{H}$, the complex Hilbert space  $\mathcal{L}(\mathcal{H})$ of all the linear operators $A\!:\!\mathcal{H}\!\rightarrow \!\mathcal{H}$ with the inner product $\langle\!\langle  A,B\rangle\!\rangle \!=\!{\rm Tr}(A^\dag B)$ and the real Hilbert space $\mathcal{A}(\mathcal{H})$ of all the self-adjoint (also called Hermitian) operators $A\!:\!\mathcal{H}\!\rightarrow \!\mathcal{H}$ with the inner product $\langle\!\langle  A,B\rangle\!\rangle \!=\!{\rm Tr}(A B)$ play a fundamental role.
Their elements can be described by using certain orthogonal bases \cite{Vourdas04,Bertlmann08,Ruzzi05,Luis98}, 
but it is known that a frame representation allows a description of the quasi-probability representations of finite quantum systems in a unified formalism \cite{Ferrie09,Ferrie11}, and has several other remarkable properties \cite{Ferrie11,DeBrota20}.

In the case when $\mathcal{H}$ is odd-dimensional, a discrete version of the Wigner function of a self-adjoint operator $A\!:\!\mathcal{H}\!\rightarrow \!\mathcal{H}$ is usually defined as $(j,k)\!\mapsto \!\frac{1}{d}{\rm Tr}(A\, \Pi(j,k))$, where $\{\Pi(j,k)\}$ is the orthogonal basis of $\mathcal{A}(\mathcal{H})$ formed by a discrete version of the displaced parity operators. 
We show that a tight frame $\{W_{jk}\}$ of $\mathcal{A}(\mathcal{H})$ can be obtained in a simple way by starting from any finite tight frame $\{|v_j\rangle\}$ of $\mathcal{H}$, and try to prove that, in certain cases, the function $(j,k)\!\mapsto \!{\rm Tr}(A\, W_{jk})$ containing all the information concerning $A$ is a useful representation. 
 The frame representations defined in this way may help us to extract some useful information concerning the quantum states of qudits, a deeper analysis of nonclassicality, coherence and entanglement of quantum states. Several examples and some possible applications are presented.

\subsection{Finite tight frames} 
Let $\mathcal{K}$ be a finite-dimensional Hilbert space over the field $\mathbb{F}$, where $\mathbb{F}\!=\!\mathbb{R}$ or $\mathbb{F}\!=\!\mathbb{C}$. In the next sections,  $\mathcal{K}\!=\!\mathcal{H}$ or $\mathcal{K}\!=\!\mathcal{L}(\mathcal{H})$
or $\mathcal{K}\!=\!\mathcal{A}(\mathcal{H})$.
A set $\{u_0,u_1,...,u_m\}$ is a  {\em spanning set} for 
$\mathcal{K}$ if any element of $\mathcal{K}$ can be represented as a linear combination of $u_0$, $u_1$,..., $u_m$ with coefficients from $\mathbb{F}$.
A sequence of vectors $u_0,\, u_1\, ,...,\, u_m$ is  a {\em frame} for $\mathcal{K}$ if there exist two constants $\alpha,\beta \!\in\!(0,\infty)$ such that
\begin{equation}\label{frame}  
\alpha||x||^2\!\leq \!\sum\limits_{k=0}^m|\langle u_k|x\rangle|^2\!\leq\! \beta||x||^2
\end{equation}
for any $x\!\in\!\mathcal{K}$.
It is known \cite{Waldron18,Christensen03} that in a finite-dimensional Hilbert space, any spanning set is a frame.\\
A frame $u_0,\, u_1,\, ...,\, u_m$ is called a {\em tight frame} if in (\ref{frame}) we can choose $\alpha \!=\!\beta$, that is if
$\sum_{k=0}^m|\langle u_k|x\rangle|^2\!\!=\! \alpha ||x||^2,$
for any $x\!\in\!\mathcal{K}$. Without loss of generality, we can assume that $\alpha\!=\!1$, prove that
$
\mathbb{I}_{\mathcal{K}}\!=\!\sum\limits_{k=0}^m |u_k\rangle \langle u_k|,
$
and consequently
\begin{equation}
\begin{array}{l}
|x\rangle \equiv\mathbb{I}_{\mathcal{K}}|x\rangle \!=\!\sum\limits_{k=0}^m |u_k\rangle \langle u_k|x\rangle ,\\
A\equiv\mathbb{I}_{\mathcal{K}}A\mathbb{I}_{\mathcal{K}} \!=\!\sum\limits_{j,k=0}^m |u_j\rangle \langle u_j|A|u_k\rangle \langle u_k| ,
\end{array}
\end{equation}
for any linear operator $A:\mathcal{K}\rightarrow\mathcal{K}$
and any $x\!\in\!\mathcal{K}$.

A tight frame can be obtained from any frame \cite{Waldron18,Christensen03}. 
If $u_0,\, u_1,\, ...,\, u_m$ is a frame, then the frame operator
\begin{equation}
S\!:\!\mathcal{K}\!\rightarrow\!\mathcal{K},\qquad  S\!=\! \sum_{k=0}^m |u_k\rangle \langle u_k|,
\end{equation}
is self-adjoint, positive-definite, invertible and
\begin{equation}
\mathbb{I}_{\mathcal{K}}\!=\!S^{-1/2}SS^{-1/2}\!=\!\sum_{k=0}^m S^{-1/2}|u_k\rangle \langle u_k|S^{-1/2},
\end{equation}
that is, $\mathbf{u}_0\!=\!S^{-1/2}u_0,\, ...,\, \mathbf{u}_m\!=\!S^{-1/2}u_m$
is a tight frame. So, a large variety of tight frames of 
$\mathcal{H}$,  $\mathcal{L}(\mathcal{H})$ and  $\mathcal{A}(\mathcal{H})$ can be obtained by starting from spanning sets of these spaces. 

Any orthonormal basis is a tight frame. A tight frame which is not an orthogonal basis contains more vectors than the dimension of the space, and the representation of a vector as a linear combination of the elements of the tight frame is not unique.
Nevertheless, the indicated standard representation is a privileged one in the sense
\begin{equation}
\left. \begin{array}{l}
|x\rangle \!=\!\sum\limits_{k=0}^m |u_k\rangle \langle u_k|x\rangle ,\\
|x\rangle \!=\!\sum\limits_{k=0}^m |u_k\rangle \, \alpha_k
\end{array}\right\}\Rightarrow
\sum\limits_{k=0}^m  |\alpha_k|^2\geq \sum\limits_{k=0}^m |\langle u_k|x\rangle |^2.
\end{equation}

For example, in the case $\mathcal{H}\!=\!\mathbb{C}^2$, the vectors
\begin{equation}
|u_0\rangle\!=\!\left( 
\begin{array}{c}
\sqrt{\frac{2}{3}}\\[2mm]
0
\end{array}\right),\quad |u_1\rangle\!=\!\left( 
\begin{array}{r}
-\frac{1}{\sqrt{6}}\\[1mm]
\frac{1}{\sqrt{2}}
\end{array}\right),\quad |u_2\rangle\!=\!\left( 
\begin{array}{r}
-\frac{1}{\sqrt{6}}\\[1mm]
-\frac{1}{\sqrt{2}}
\end{array}\right),
\end{equation}
satisfying the relation $|u_0\rangle\!+\!|u_1\rangle\!+\!|u_2\rangle\!=\!0$, form a tight frame, and for any $|x\rangle\!\in \!\mathbb{C}^2$ and any $\lambda\!\in \!\mathbb{C}$, we have
\begin{equation}
\begin{array}{l}
|x\rangle \!=\!\sum\limits_{k=0}^2 |u_k\rangle \langle u_k|x\rangle \!=\!\sum\limits_{k=0}^2 |u_k\rangle (\langle u_k|x\rangle \!+\!\lambda),\\
\sum\limits_{k=0}^2 |\langle u_k|x\rangle \!+\!\lambda|^2\!=\!\sum\limits_{k=0}^2 |\langle u_k|x\rangle|^2 \!+\!|\lambda|^2\!\geq\!\sum\limits_{k=0}^2 |\langle u_k|x\rangle|^2.
\end{array}.
\end{equation}

The orthogonal projection of an orthonormal basis of a Hilbert space onto a subspace is a tight frame in  subspace.
Any tight frame $\{u_0,u_1,...,u_m\}$ of $\mathcal{H}$ allows us to embed $\mathcal{H}$ into the Hilbert space $\mathbb{F}^{m+1}$ such that $\{u_0,u_1,...,u_m\}$ is the orthogonal projection of the canonical basis $\{\varepsilon_0,\varepsilon_1,...,\varepsilon_m\}$ of  $\mathbb{F}^{m+1}$. If $\{e_0,e_1,...,e_{d-1}\}$
is an orthonormal basis of $\mathcal{H}$, then
$\{w_0,w_1,...,w_{d-1}\}$, where
\begin{equation}
|w_k\rangle \!=\!\left( 
\begin{array}{c}
\langle u_0|e_k\rangle\\
\langle u_1|e_k\rangle\\
\vdots\\
\langle u_m|e_k\rangle\\
\end{array}\right), 
\end{equation}
is an orthonormal system,
\[
\langle w_j|w_k\rangle \!=\!\sum\limits_{i=0}^m\overline{\langle u_i|e_j\rangle}\langle u_i|e_k\rangle
\!=\!\sum\limits_{i=0}^m\langle e_j|u_i\rangle\langle u_i|e_k\rangle \!=\!\langle e_j|e_k\rangle\!=\!\delta_{jk}.
\]
The Hilbert space $\mathcal{H}$ can be identified with the subspace $\mathcal{S}\!=\!\mathrm{span}\{w_0,w_1,...,w_{d-1}\}$ of $\mathbb{F}^{m+1}$ by using the linear isometry 
\begin{equation}\label{embedding}
\mathcal{H}\!\rightarrow\!\mathcal{S}:\quad \sum\limits_{j=0}^{d-1}\alpha_j|e_j\rangle \ \mapsto \ \sum\limits_{j=0}^{d-1}\alpha_j|w_j\rangle .
\end{equation}
The orthogonal projector corresponding to $\mathcal{S}$ is
$P\!:\!\mathbb{F}^{m+1}\!\rightarrow\!\mathcal{S}$, \ 
$P\!=\!\sum\limits_{j=0}^{d-1}|w_j\rangle\langle w_j|$, and
\[
\begin{array}{r}
P|\varepsilon_k\rangle\!=\!\sum\limits_{j=0}^{d-1}|w_j\rangle\langle w_j|\varepsilon_k\rangle\!=\!\sum\limits_{j=0}^{d-1}|w_j\rangle\overline{\langle\varepsilon_k|w_j\rangle}\!=\!\sum\limits_{j=0}^{d-1}|w_j\rangle\overline{\langle u_k|w_j\rangle}\\
=\!\sum\limits_{j=0}^{d-1}|w_j\rangle\langle w_j|u_k\rangle\!\equiv\!\sum\limits_{j=0}^{d-1}|e_j\rangle\langle e_j|u_k\rangle\!=\!|u_k\rangle.
\end{array}
\]

\subsection{A finite version of the Wigner function}
The finite versions of Wigner function \cite{Vourdas04,Ferrie11,Wootters87,Feynman87,Leonhardt95,Leonhardt96,Opatrny95,Galetti96,Cotfas23} play an important role in description of qudits.
For simplicity, in this section,  we consider only quantum systems with 
an odd-dimensional ($d\!=\!2s\!+\!1$) Hilbert space $\mathcal{H}$, which can be regarded as the space of all the functions 
$\psi\!:\!\{-\!s,\!-s\!+\!1,...,s\!-\!1,s\}\!\rightarrow \!\mathbb{C}$
with the inner product
$
\langle \varphi|\psi\rangle \!=\!\sum_{j=-s}^s\overline{\varphi(j)}\,\psi(j).
$
Each function $\psi\!\in\!\mathcal{H}$ can be extended up to a periodic function $\psi\!:\!\mathbb{Z}\!\rightarrow \!\mathbb{C}$
of period $d$. 

The {\em displacement operators} 
$D(j ,k )\!:\!\mathcal{H}\!\rightarrow\!\mathcal{H},$
\begin{equation}
 D(j ,k )\psi (n)={\rm e}^{-(\pi {\rm i}/d)kj}\,{\rm e}^{(2\pi {\rm i}/d)kn}\, \psi (n\!-\!j ),
\end{equation}
form \cite{Vourdas04,Opatrny95,Galetti96,Cotfas23}  an orthogonal basis in $\mathcal{L}(\mathcal{H})$, 
\begin{equation}
\begin{array}{c}
\langle\!\langle D(j,k), D(n,m)\rangle\!\rangle\!=\!d\, \delta_{jn}\, \delta_{km},\\[2mm]
\mathbb{I}_{\mathcal{L}(\mathcal{H})}\!=\!\frac{1}{d}\sum\limits_{j,k=-s}^s|D(j,k)\rangle\!\rangle\langle\!\langle D(j,k)|,
\end{array}
\end{equation}
and the {\em displaced parity operators} 
$\Pi(j,k)\!:\!\mathcal{H}\!\rightarrow\!\mathcal{H},$
\begin{equation}\label{Pijk}
\Pi (j ,k )= D(j,k)\, \Pi\,  D^\dag (j,k),
\end{equation}
where $\Pi \!:\!\mathcal{H}\!\rightarrow\!\mathcal{H},\ \Pi \psi(n)\!=\!\psi(-n) $, form \cite{Vourdas04,Opatrny95,Galetti96,Cotfas23} an orthogonal basis in 
$\mathcal{A}(\mathcal{H})$,
\begin{equation}
\begin{array}{l}
\langle\!\langle \Pi (j,k), \Pi(n,m)\rangle\!\rangle\!=\!d\, \delta_{jn}\, \delta_{km},\\[2mm] 
\mathbb{I}_{\mathcal{A}(\mathcal{H})}\!=\!\frac{1}{d}\sum\limits_{j,k=-s}^s|\Pi(j,k)\rangle\!\rangle\langle\!\langle \Pi(j,k)|.
\end{array}
\end{equation}

Any operator $A\!\in\!\mathcal{A}(\mathcal{H})$ admits the representations
\begin{equation}
A\!=\!\!\sum\limits_{j,k=-s}^s\!\!\!\chi_{{}_A}(j,k)\,D(j,k)\!=\!\!\sum_{j,k=-s}^s\!\!\mathcal{W}_A(j,k)\,\Pi(j,k),
\end{equation}
where 
\begin{equation}
\begin{array}{r}
\chi_{{}_A}\!:\!\{-\!s,\!-s\!+\!1,...,s\}\!\times\!\{-\!s,\!-s\!+\!1,...,s\}\!\longrightarrow\!\mathbb{C},\\
\chi_{{}_A}(j,k)\!=\!\frac{1}{d}\langle\!\langle D(j,k)|A\rangle\!\rangle \!=\!\frac{1}{d}\mathrm{Tr}(A\, D^\dag(j,k))
\end{array}
\end{equation}
is the {\em Weyl characteristic function} \cite{Vourdas04,Leonhardt95,Leonhardt96} of $A$, and
\begin{equation}
\begin{array}{r}
\mathcal{W}_A\!:\!\{-\!s,\!-s\!+\!1,...,s\}\!\times\!\{-\!s,\!-s\!+\!1,...,s\}\!\longrightarrow\!\mathbb{R},\\
\mathcal{W}_A(j,k)\!=\!\frac{1}{d}\langle\!\langle \Pi(j,k)|A\rangle\!\rangle \!=\!\frac{1}{d}\mathrm{Tr}(A\, \Pi(j,k)),
\end{array}
\end{equation}
is the {\em Wigner function} \cite{Vourdas04,Cotfas23} of $A$. If $A,B\!\in\!\mathcal{A}(\mathcal{H})$, then
\begin{equation}
\begin{array}{c}
\mathrm{Tr}(AB)\!=\!\langle\!\langle A|B\rangle\!\rangle\!=\!\frac{1}{d}\!\!\!\sum\limits_{j,k=-s}^s\!\!\langle\!\langle A|D(j,k)\rangle\!\rangle\langle\!\langle D(j,k)|B\rangle\!\rangle,\\
\mathrm{Tr}(AB)\!=\!\langle\!\langle A|B\rangle\!\rangle\!=\!\frac{1}{d}\!\!\!\sum\limits_{j,k=-s}^s\!\!\langle\!\langle A|\Pi(j,k)\rangle\!\rangle\langle\!\langle \Pi(j,k)|B\rangle\!\rangle,
\end{array}
\end{equation}
that is
\begin{equation}
\mathrm{Tr}(AB)\!=\!d\!\sum\limits_{j,k=-s}^s\!\!\!\overline{\chi_{{}_A}(j,k)}\,\chi_{{}_B}(j,k)\!=\!d\!\sum\limits_{j,k=-s}^s\!\!\!\mathcal{W}_A(j,k)\,\mathcal{W}_B(j,k).
\end{equation}

For any $\kappa \!\in\!(0,\infty)$, the function
\begin{equation}
g_\kappa \!:\!\{-\!s,\!-s\!+\!1,...,s\!-\!1,s\}\!\rightarrow\!\mathbb{R},\quad 
g_\kappa(n)\!=\!\sum\limits_{m=-\infty}^\infty \mathrm{e}^{-(\kappa \pi/d)(n+md)^2}
\end{equation}
represents a discrete version of the Gaussian function 
$\mathbb{R}\!\rightarrow\!\mathbb{R}:q\mapsto \mathrm{e}^{-(\kappa \pi/h)q^2}$. Under the Fourier transform
\begin{equation}
\mathcal{F}\!:\!\mathcal{H}\!\rightarrow\!\mathcal{H}\!:\!\psi\mapsto \mathcal{F}[\psi],\quad 
\mathcal{F}[\psi](k)\!=\!\frac{1}{\sqrt{d}}\sum\limits_{n=-s}^s \mathrm{e}^{-(2\pi\mathrm{i}/d)kn}\psi(n),
\end{equation}
the discrete Gaussian function transforms as
\begin{equation}
\mathcal{F}[g_\kappa]\!=\!\frac{1}{\sqrt{\kappa}}\, g_{1/\kappa}.
\end{equation}
Particularly, the normalized function
\begin{equation}
|0,0\rangle \!=\!\frac{1}{||g_1||}\, |g_1\rangle
\end{equation}
is an eigenfunction of $\mathcal{F}$ corresponding to the eigenvalue 1, and can be regarded as a discrete version of the vacuum state.\\ The discrete coherent states
\begin{equation}
|j,k\rangle \!=\!D(j,k)|0,0\rangle
\end{equation}
form a tight frame in $\mathcal{H},$
\begin{equation}
\frac{1}{d}\sum\limits_{j,k=-s}^s|j,k\rangle \langle j,k|\!=\!\mathbb{I}_\mathcal{H}.
\end{equation}

\subsection{A discrete phase-space representation of qutrit}
In the case of the qutrit, $\mathbb{C}^3\equiv\{\psi\!:\!\{-1,0,1\}\!\rightarrow \!\mathbb{C}\}$,
and the displacement operators are
  {\footnotesize 
\begin{equation}\fl
\begin{array}{rrr}
D(-1,-1 )\!=\!\left(\!\!
\begin{array}{ccc}
0 & {\rm e}^{\frac{\pi {\rm i}}{3}} & 0\\
0 & 0 &  {\rm e}^{-\frac{\pi {\rm i}}{3}}\\
-1 & 0 & 0
\end{array}\!\!\right), &
D(-1,0)\!=\!\left(\!\!
\begin{array}{ccc}
0 & 1 & 0\\
0 & 0 & 1\\
1 & 0 & 0
\end{array}\!\!\right), &
D(-1,1 )\!=\!\left(\!\!
\begin{array}{ccc}
0 & {\rm e}^{-\frac{\pi {\rm i}}{3}} & 0\\
0 & 0 &  {\rm e}^{\frac{\pi {\rm i}}{3}}\\
-1 & 0 & 0
\end{array}\!\!\right),\\[8mm]
D(0,-1 )\!=\!\left(\!\!
\begin{array}{ccc}
{\rm e}^{\frac{2\pi {\rm i}}{3}} & 0 & 0\\
0 & 1 &  0\\
0 & 0 & {\rm e}^{-\frac{2\pi {\rm i}}{3}}
\end{array}\!\!\right), &
D(0,0)\!=\!\left(\!\!
\begin{array}{ccc}
1 & 0 & 0\\
0 & 1 & 0\\
0 & 0 & 1
\end{array}\!\!\right), &
D(0,1 )\!=\!\left(\!\!
\begin{array}{ccc}
{\rm e}^{-\frac{2\pi {\rm i}}{3}} & 0 & 0\\
0 & 1 &  0\\
0 & 0 & {\rm e}^{\frac{2\pi {\rm i}}{3}}
\end{array}\!\!\right),\\[8mm]
D(1,-1 )\!=\!\left(\!\!
\begin{array}{ccc}
0 & 0 & -1\\
{\rm e}^{\frac{\pi {\rm i}}{3}} & 0 &  0\\
0 & {\rm e}^{-\frac{\pi {\rm i}}{3}} & 0
\end{array}\!\!\right), &
D(1,0)\!=\!\left(\!\!
\begin{array}{ccc}
0 & 0 & 1\\
1 & 0 & 0\\
0 & 1 & 0
\end{array}\!\!\right), &
D(1,1 )\!=\!\left(\!\!
\begin{array}{ccc}
0 & 0 & -1\\
{\rm e}^{-\frac{\pi {\rm i}}{3}} & 0 &  0\\
0 & {\rm e}^{\frac{\pi {\rm i}}{3}} & 0
\end{array}\!\!\right).
\end{array}
\end{equation}
}
The Wigner function of a linear operator $A\!\in\!\mathcal{A}(\mathbb{C}^3)$ is
\begin{equation}\fl
\mathcal{W}_A\!:\!\{-1,0,1\}\!\times\!\{-1,0,1\}\!\longrightarrow\!\mathbb{R},\qquad
\mathcal{W}_A(j,k)\!=\!\frac{1}{3}\langle\!\langle \Pi(j,k)|A\rangle\!\rangle \!=\!\frac{1}{3}\mathrm{Tr}(A\, \Pi(j,k)),
\end{equation}
where the displaced parity operators  $\Pi(j,k)$ are
  {\footnotesize 
\begin{equation}\fl
 \begin{array}{rrr}
\Pi(-1,-1 )\!=\!\left(\!\!
\begin{array}{ccc}
1 & 0 & 0\\
0 & 0 &  -{\rm e}^{-\frac{\pi {\rm i}}{3}}\\
0 & -{\rm e}^{\frac{\pi {\rm i}}{3}} & 0
\end{array}\!\!\right), &
\Pi(-1,0)\!=\!\left(\!\!
\begin{array}{ccc}
1 & 0 & 0\\
0 & 0 & 1\\
0 & 1 & 0
\end{array}\!\!\right), &
\Pi(-1,1 )\!=\!\left(\!\!
\begin{array}{ccc}
1 & 0 & 0\\
0 & 0 &  -{\rm e}^{\frac{\pi {\rm i}}{3}}\\
0 & -{\rm e}^{-\frac{\pi {\rm i}}{3}} & 0
\end{array}\!\!\right),\\[8mm]
\Pi(0,-1 )\!=\!\left(\!\!
\begin{array}{ccc}
0 & 0 & {\rm e}^{-\frac{2\pi {\rm i}}{3}}\\
0 & 1 &  0\\
{\rm e}^{\frac{2\pi {\rm i}}{3}} & 0 & 0
\end{array}\!\!\right), &
\Pi(0,0)\!=\!\left(\!\!
\begin{array}{ccc}
0 & 0 & 1\\
0 & 1 & 0\\
1 & 0 & 0
\end{array}\!\!\right), &
\Pi(0,1 )\!=\!\left(\!\!
\begin{array}{ccc}
0 & 0 & {\rm e}^{\frac{2\pi {\rm i}}{3}}\\
0 & 1 &  0\\
{\rm e}^{-\frac{2\pi {\rm i}}{3}} & 0 & 0
\end{array}\!\!\right),\\[8mm]
\Pi(1,-1 )\!=\!\left(\!\!
\begin{array}{ccc}
0 & -{\rm e}^{-\frac{\pi {\rm i}}{3}} & 0\\
-{\rm e}^{\frac{\pi {\rm i}}{3}} & 0 &  0\\
0 & 0 & 1
\end{array}\!\!\right), &
\Pi(1,0)\!=\!\left(\!\!
\begin{array}{ccc}
0 & 1 & 0\\
1 & 0 & 0\\
0 & 0 & 1
\end{array}\!\!\right), &
\Pi(1,1 )\!=\!\left(\!\!
\begin{array}{ccc}
0 & -{\rm e}^{\frac{\pi {\rm i}}{3}} & 0\\
-{\rm e}^{-\frac{\pi {\rm i}}{3}} & 0 &  0\\
0 & 0 & 1
\end{array}\!\!\right).
\end{array}
\end{equation}
}
The discrete Fourier transform
\begin{equation}
\mathcal{F}\!:\!\mathbb{C}^3\!\longrightarrow\!\mathbb{C}^3,\qquad \mathcal{F}\!=\!\frac{1}{\sqrt{3}}\left(\!\!
\begin{array}{ccc}
{\rm e}^{-\frac{2\pi {\rm i}}{3}} & 1 & {\rm e}^{\frac{2\pi {\rm i}}{3}}\\
1 & 1 & 1\\
{\rm e}^{\frac{2\pi {\rm i}}{3}} & 1 & {\rm e}^{-\frac{2\pi {\rm i}}{3}}
\end{array}\!\!\right)
\end{equation}
 has the distinct eigenvalues $1$, $-{\rm i}$, $-1$, and the spectral decomposition
 \begin{equation}
\mathcal{F}\!=\!|\mathfrak{F}_0\rangle\langle \mathfrak{F}_0|\!-{\rm i}
|\mathfrak{F}_{1}\rangle\langle \mathfrak{F}_{1}|\!-\!
|\mathfrak{F}_{2}\rangle\langle \mathfrak{F}_{2}|,
\end{equation}
where
\begin{equation}
|\mathfrak{F}_0\rangle\!\!=\!\!\left(\!\!
\begin{array}{c}
\frac{1}{2}\sqrt{1-\frac{1}{\sqrt{3}}}\\
\frac{1}{\sqrt{2}}\sqrt{1+\frac{1}{\sqrt{3}}}\\
\frac{1}{2}\sqrt{1-\frac{1}{\sqrt{3}}}
\end{array}\!\!\right)\!\!, \ \ 
|\mathfrak{F}_1\rangle\!=\!\!
\left(\!\!\!
\begin{array}{c}
-\frac{1}{\sqrt{2}}\\
0\\
\frac{1}{\sqrt{2}}
\end{array}\!\!\!\right)\!\!,\ \ 
|\mathfrak{F}_2\rangle\!\!=\!\!\left(\!\!\!
\begin{array}{c}
\frac{1}{2}\sqrt{1+\frac{1}{\sqrt{3}}}\\
-\frac{1}{\sqrt{2}}\sqrt{1-\frac{1}{\sqrt{3}}}\\
\frac{1}{2}\sqrt{1+\frac{1}{\sqrt{3}}}
\end{array}\!\!\!\right).
\end{equation}
The discrete vacuum state is $|0,0\rangle\!=\!|\mathfrak{F}_0\rangle$, and the nine discrete coherent states $|j,k\rangle \!=\!D(j,k)|0,0\rangle$
form a tight frame in $\mathbb{C}^3$, namely $\frac{1}{3}\sum\limits_{j,k=-1}^1|j,k\rangle \langle j,k|\!=\!\mathbb{I}_{\mathbb{C}^3}$.

\subsection{A discrete phase-space representation of qubit}
In the case of the qubit, the operators
 \begin{equation}
\begin{array}{ll}
K_{00}\!=\!\frac{1}{2}(\mathbb{I}\!+\!\sigma_x\!+\!\sigma_y\!+\!\sigma_z),\qquad {} & 
K_{01}\!=\!\frac{1}{2}(\mathbb{I}\!-\!\sigma_x\!-\!\sigma_y\!+\!\sigma_z),\\[2mm]
K_{10}\!=\!\frac{1}{2}(\mathbb{I}\!+\!\sigma_x\!-\!\sigma_y\!-\!\sigma_z), & 
K_{11}\!=\!\frac{1}{2}(\mathbb{I}\!-\!\sigma_x\!+\!\sigma_y\!-\!\sigma_z)
\end{array}
\end{equation}
proposed independently by Wootters \cite{Wootters87}  and Feynman \cite{Feynman87}, form an orthogonal basis in $\mathcal{A}(\mathbb{C}^2).$
For $A\!\in\!\mathcal{A}(\mathbb{C}^2)$, the function
\begin{equation}
\mathcal{W}_A\!:\!\{0,1\}\!\times\!\{0,1\}\!\longrightarrow\!\mathbb{R},\qquad
\mathcal{W}_A(j,k)\!=\!\frac{1}{2}\langle\!\langle K_{jk}|A\rangle\!\rangle \!=\!\frac{1}{2}\mathrm{Tr}(A\, K_{jk}),
\end{equation}
represents the Wigner function of $A$.

\subsection{A continuous phase-space representation of qubit}
In the case of a two-level quantum system (qubit), the surface of a sphere is used as a phase space, and the continuous Wigner function of a state $\varrho$ is defined as \cite{Ferrie11,Tilma16,Rundle17,Koczor20,Rundle21}
\begin{equation}
\mathcal{W}_\varrho(\theta,\phi)\!=\!\mathrm{Tr}(\varrho\, \Delta(\theta,\phi)),
\end{equation}
where $\theta\!\in\![0,\pi]$, $\phi\!\in\![0,2\pi)$ are the Euler angles,
\begin{equation}\label{RPiR}
\Delta(\theta,\phi)\!=\!R(\theta,\phi,\Phi)\, \Pi\, R^\dag(\theta,\phi,\Phi),
\end{equation}
$\Pi\!=\!\frac{1}{2}(\mathbb{I}\!+\!\sqrt{3}\, \sigma _z)$, and $R(\theta,\phi,\Phi)$ is the rotation operator $R(\theta,\phi,\Phi)\!=\!{\rm e}^{-{\rm i}\phi \sigma_z/2}{\rm e}^{-{\rm i}\theta \sigma_y/2}{\rm e}^{-{\rm i}\Phi \sigma_z/2}$ .

In the case of a composite system of $N$ qubits, the phase space is a product of $N$ spheres, and the Wigner function of a state $\varrho$ is defined as
\begin{equation}
\mathcal{W}_\varrho(\theta_1,...,\theta_N,\phi_1,...,\phi_N)\!=\!\mathrm{Tr}\left(\varrho\, \bigotimes_{i=1}^N\Delta(\theta_i,\phi_i)\right).
\end{equation}
For visualization of a state of the composite system, it is often sufficient to consider the equal-angle slice \cite{Tilma16}
\begin{equation}
\mathcal{W}^{EA}_\varrho(\theta,\phi)\!=\!\mathcal{W}_\varrho(\theta,...,\theta,\phi,...,\phi).
\end{equation}
\section{Frame representations of qudit quantum states and observables}
\subsection{Wigner function of a single qudit}
In order to obtain a frame representation for a qudit with the complex Hilbert space $\mathcal{H}$ of dimension $d$, 
we start from a tight frame $v_0,\, v_1,\, ...,\, v_r $ for $\mathcal{H}$ satisfying 
$\mathbb{I}_{\mathcal{H}}\!=\!\sum_{j=0}^r  |v_j\rangle \langle v_j|.$\\
For any $A\!\in\!\mathcal{L}(\mathcal{H})$, we have
\begin{equation}
A \!=\!\mathbb{I}_{\mathcal{H}}\,A\,\mathbb{I}_{\mathcal{H}} \!=\!\sum\limits_{j,k=0}^r  |v_j\rangle \langle v_j|A|v_k\rangle \langle v_k|
\!=\!\sum\limits_{j,k=0}^r  \langle v_j|A|v_k\rangle |v_j\rangle \langle v_k|.
\end{equation}
Particularly, this means that the operators
\begin{equation}
V_{jk}\!=\!|v_j\rangle \langle v_k|, \qquad \mbox{where} \ j,k\!\in\!\{0,1,...,r\}
\end{equation}
span $\mathcal{L}(\mathcal{H})$, and consequently  form a frame. \\[3mm]
{\bf Theorem 1.} {\em The frame $\{ V_{jk}\}$ is a tight frame for the complex Hilbert space $\mathcal{L}(\mathcal{H})$, that is}
\begin{equation}
\mathbb{I}_{\mathcal{L}(\mathcal{H})}\!=\!\sum_{j,k=0}^r |V_{jk}\rangle\!\rangle\langle \!\langle V_{jk}|.
\end{equation}
{\em Proof.} If $\{|e_j\rangle \}$ is an orthonormal basis of $\mathcal{H}$, then  $\{|E_{jk}\rangle\!\rangle\!=\!|e_j\rangle\langle e_k|\}$ is an orthonormal basis of $\mathcal{L}(\mathcal{H})$. Since
\begin{equation}
\mathbb{I}_{\mathcal{H}}\!=\!\sum_{j=0}^r  |v_j\rangle \langle v_j|\quad \Rightarrow \quad 
\sum_{j=0}^r  \langle e_n|v_j\rangle \langle v_j|e_m\rangle \!=\!\delta_{nm},
\end{equation}
$(|u\rangle\langle v|)^\dag\!=\!|v\rangle\langle u|$, and $\mathrm{Tr}(|e_m\rangle \langle e_n| A)\!=\!\langle e_n|A|e_m\rangle$, the frame operator \ $S\!=\!\sum_{j,k=0}^r |V_{jk}\rangle\!\rangle\langle \!\langle V_{jk}|$ satisfies
\[
\begin{array}{l}
\langle\!\langle E_{nm}|S|E_{i\ell}\rangle\!\rangle \!=\!\sum_{j,k=0}^r \langle\!\langle E_{nm}|V_{jk}\rangle\!\rangle\langle \!\langle V_{jk}|E_{i\ell}\rangle\!\rangle \\[1mm]
\quad =\!\sum_{j,k=0}^r \mathrm{Tr}(|e_m\rangle \langle e_n|\ |v_j\rangle \langle v_k|) \ \mathrm{Tr}(|v_k\rangle \langle v_j|\ |e_i\rangle \langle e_\ell|\ )\\[1mm]
\quad =\!\sum_{j,k=0}^r \langle e_n|v_j\rangle \langle v_k|e_m\rangle \ \langle v_j|e_i\rangle \langle e_\ell|v_k\rangle \\[1mm]
\quad =\!\sum_{j=0}^r \langle e_n|v_j\rangle \langle v_j|e_i\rangle \sum_{k=0}^r 
\langle e_\ell|v_k\rangle \langle v_k|e_m\rangle 
\!=\!\delta_{ni}\delta_{m\ell}.\ \rule{2mm}{2mm}
\end{array}
\]

Any operator $A\!\in\!\mathcal{L}(\mathcal{H})$ admits the representation
\begin{equation}
A\!=\!\sum\limits_{j,k=0}^r  \chi_{{}_A}(j,k)\,V_{jk},\quad \mathrm{where}\quad
\begin{array}{l}
\chi_{{}_A}\!:\!\{0,1,...,r \}\!\times\!\{0,1,...,r \}\!\rightarrow\!\mathbb{C},\\[2mm]
\chi_{{}_A}(j,k)\!=\!\langle \!\langle V_{jk}|A\rangle\!\rangle \!=\!\mathrm{Tr}(A\,V_{jk}^\dag).
\end{array}
\end{equation}
It can be regarded as a version of the {\em discrete Weyl characteristic function of} $A$. The displacement operators $D(j,k)$ are unitary operators
satisfying the relation 
\begin{equation}\label{DjkDnm}
D(j,k) D(n,m)\!=\!\mathrm{e}^{(\pi \mathrm{i}/d)(kn-jm)}D(j\!+\!n,k\!+\!m).
\end{equation}
The operators $V_{jk}$ satisfy $V_{jk}^\dag\!=\!V_{kj}$ and the relations
\begin{equation}
V_{jk}\, V_{nm}\!=\!\langle v_k|v_n\rangle  V_{jm},
\end{equation}
instead of (\ref{DjkDnm}), but they are not unitary operators.\\[3mm]
{\bf Theorem 2.}  {\em In the real Hilbert space $\mathcal{A}(\mathcal{H})$,  the  self-adjoint operators 
\begin{equation}
\begin{array}{lll}
W_{jj}\!=\!V_{jj},\quad & \mbox{for}\quad & 0\leq j\leq r\\[2mm]
W_{jk}\!=\!\frac{1}{\sqrt{2}}(V_{jk}\!+\!V_{kj}),\quad & \mbox{for}\quad & 0\leq j<k\leq r\\[2mm]
W_{jk}\!=\!\frac{\rm i}{\sqrt{2}}(V_{kj}\!-\!V_{jk}),\quad & \mbox{for}\quad & 0\leq k<j\leq r
\end{array}
\end{equation}
form a tight frame, that is} 
$
\mathbb{I}_{\mathcal{A}(\mathcal{H})}\!=\!\sum_{j,k=0}^r |W_{jk}\rangle\!\rangle\langle \!\langle W_{jk}|.\\
$
{\em Proof.} In $\mathcal{L}(\mathcal{H})$, \  $\mathbf{S}\!=\!\sum_{j,k=0}^r |W_{jk}\rangle\!\rangle\langle \!\langle W_{jk}|$ satisfies
\begin{equation}
\mathbf{S}\!=\!\sum_{j,k=0}^r |W_{jk}\rangle\!\rangle\langle \!\langle W_{jk}|\!=\!\sum_{j,k=0}^r |V_{jk}\rangle\!\rangle\langle \!\langle V_{jk}|\!=\!\mathbb{I}_{\mathcal{L}(\mathcal{H})},
\end{equation}
but the restriction $\mathbf{S}|_{\mathcal{A}(\mathcal{H})} $ of $\mathbf{S}$ to $\mathcal{A}(\mathcal{H})$ is $\mathbb{I}_{\mathcal{A}(\mathcal{H})}.\quad \rule{2mm}{2mm}$\\[3mm]
Any operator $A\!\in\!\mathcal{A}(\mathcal{H})$ admits the representation
\begin{equation}
A\!=\!\sum\limits_{j,k=0}^r \mathcal{W}_{{}_A}(j,k)\,W_{jk},
\quad \mathrm{where}\quad
\begin{array}{l}
\mathcal{W}_{{}_A}\!:\!\{0,1,...,r\}\!\times\!\{0,1,...,r\}\!\rightarrow\!\mathbb{R},\\[2mm]
\mathcal{W}_{{}_A}(j,k)\!=\!\langle \!\langle W_{jk}|A\rangle\!\rangle\!=\!\mathrm{Tr}(A\,W_{jk}).
\end{array}
\end{equation}
It can be regarded as a version of the discrete {\em  Wigner function of} $A$. Explicitely,
\begin{equation}
\mathcal{W}_{{}_A}(j,k)\!=\!\left\{
\begin{array}{lll}
\langle v_j|A|v_j\rangle ,\quad & \mbox{for}\quad & k\!=\!j,\\[2mm]
\sqrt{2}\, \mathfrak{Re}\langle v_k|A|v_j\rangle ,\quad & \mbox{for}\quad & k\!>\!j,\\[2mm]
\sqrt{2}\, \mathfrak{Im}\langle v_k|A|v_j\rangle ,\quad & \mbox{for}\quad & k\!<\!j.
\end{array}\right.
\end{equation}

The representation of an operator
$A\!\in\!\mathcal{A}(\mathcal{H})$ as a linear combination of $W_{jk}$ is not unique, but the representation corresponding to $\mathcal{W}_{{}_A}(j,k)$ is a privileged one satisfying the condition of extremum
\begin{equation}
A\!=\!\sum\limits_{j,k=0}^r \alpha_{jk}\,W_{jk}\quad \Rightarrow \quad \sum\limits_{j,k=0}^r |\alpha_{jk}|^2\geq
\sum\limits_{j,k=0}^r \mathcal{W}_{{}_A}^2(j,k).
\end{equation}

If $A,B\!\in\!\mathcal{A}(\mathcal{H})$, then 
$\mathrm{Tr}(AB)\!=\!\langle\!\langle A|\mathbb{I}_{\mathcal{A}(\mathcal{H})}|B\rangle\!\rangle$, that is
\begin{equation}
\mathrm{Tr}(AB)\!=\!\!\!\sum\limits_{j,k=0}^r\!\overline{\chi_{{}_A}(j,k)}\,\chi_{{}_B}(j,k)\!=\!\!\!\!\sum\limits_{j,k=-0}^r\!\!\!\mathcal{W}_A(j,k)\,\mathcal{W}_B(j,k).
\end{equation}

If $U\!:\!\mathcal{H}\rightarrow\mathcal{H}$ is a unitary operator, then $Uv_0,\, Uv_1,\, ...,\, Uv_r $
is also a tight frame for $\mathcal{H}$, $\{W'_{jk}\!=\!UW_{jk}U^\dag\}$
is a tight frame for $\mathcal{A}(\mathcal{H})$, and
\begin{equation}
\mathcal{W}_{U^\dag AU}(j,k)\!=\!\mathrm{Tr}(U^\dag AU\,W_{jk})\!=\!\mathrm{Tr}(A\,W'_{jk}).
\end{equation}
The kernel $W_{jk}$, which generate $\mathcal{W}_{{}_A}(j,k)$ according to the generalized Weyl rule $\mathcal{W}_{{}_A}(j,k)\!=\!\mathrm{Tr}(A\,W_{jk})$,
satisfies the restricted version of the Stratonovich-Weyl correspondence presented in \cite{Tilma16,Brif99} except the standardization. Generally,  $\sum_{j,k=0}^r W_{jk}\!\neq\!\mathbb{I}_{\mathcal{H}}$, and consequently
$\sum_{j,k=0}^r\mathcal{W}_{{}_A}(j,k)$ depends on $A$, and not only on $\mathrm{Tr}\, A$.

If we start from an orthonormal basis $\{|v_j\rangle\}$ of $\mathcal{H}$, then $\{ V_{jk}\}$ is an orthonormal basis in 
$\mathcal{L}(\mathcal{H})$ and $\{ W_{jk}\}$ an orthonormal basis in $\mathcal{A}(\mathcal{H})$.
For example, in the case $\mathcal{H}\!=\!\mathbb{C}^2$, by starting from $\left\{|v_0\rangle\!=\!\left(\! 
\begin{array}{c}
1\\
0
\end{array}\!\right),\ \{|v_1\rangle\!=\!\left( \!
\begin{array}{c}
0\\
1
\end{array}\!\right) \right\}$ we get
{\footnotesize 
\begin{equation}
V_{00}\!=\!\!\left(\!
\begin{array}{cc}
1 & 0 \\[2mm]
0 & 0
\end{array}\!\right),\quad 
V_{01}\!=\!\!\left(\!
\begin{array}{cc}
0 & 1 \\[2mm]
0 & 0
\end{array}\!\right),\quad V_{10}\!=\!\!\left(\!
\begin{array}{cc}
0 & 0 \\[2mm]
1 & 0
\end{array}\!\right),\quad 
V_{11}\!=\!\!\left(\!
\begin{array}{cc}
0 & 0 \\[2mm]
0 & 1
\end{array}\!\right)
\end{equation}
}
and
{\footnotesize 
\begin{equation}
W_{00}\!=\!\!\left(\!\!
\begin{array}{cc}
1 & 0 \\[2mm]
0 & 0
\end{array}\!\!\right)\!,\quad 
W_{01}\!=\!\!\frac{1}{\sqrt{2}}\left(\!\!
\begin{array}{cc}
0 & 1 \\[2mm]
1 & 0
\end{array}\!\!\right)\!,\quad W_{10}\!=\!\!\frac{1}{\sqrt{2}}\left(\!\!
\begin{array}{cc}
0 & \mathrm{i} \\[2mm]
-\mathrm{i} & 0
\end{array}\!\!\right)\!,\quad 
W_{11}\!=\!\!\left(\!\!
\begin{array}{cc}
0 & 0 \\[2mm]
0 & 1
\end{array}\!\!\right)\!.
\end{equation}
}

In the case of a frame representation,
\begin{equation}
\sum_{j=0}^r  |v_j\rangle \langle v_j|\!=\!\mathbb{I}_{\mathcal{H}}\quad \Rightarrow \quad 
\sum_{j=0}^r  W_{jj}\!=\!\mathbb{I}_{\mathcal{H}},
\end{equation}
and consequently
\begin{equation}
\sum_{j=0}^r  \mathcal{W}_A(j,j)\!=\!\sum_{j=0}^r \mathrm{Tr}(A\, W_{jj})\!=\!\mathrm{Tr}\,A,
\end{equation}
for any $A\!\in\!\mathcal{A}(\mathcal{H})$. In the case of a quantum state $\varrho$,
\begin{equation}
\sum_{j=0}^r  \mathcal{W}_\varrho(j,j)\!=\!1,
\end{equation}
the {\em purity} of   $\varrho$ is
\begin{equation}
\mathrm{Tr}\,\varrho^2\!=\!\mathrm{Tr}(\varrho\,\varrho)\!=\!\sum_{j,k=0}^r  \mathcal{W}^2_\varrho(j,k),
\end{equation}
and
\begin{equation}
|\mathcal{W}_\varrho(j,k)|\!\leq\!\sqrt{\mathrm{Tr}\,\varrho^2}\!\leq\!1.
\end{equation}
\subsection{Wigner function of a composite system}
Let $\{ v_0^1,\, v_1^1,..., v_{r_1}^1 \}$ and $\{ v_0^2,\, v_1^2,..., v_{r_2}^2 \}$ be tight frames in 
the Hilbert spaces $\mathcal{H}_1$ and $\mathcal{H}_2$, and let $\{W_{jk}^1\}$ and  $\{W_{\ell m}^2\}$ be
the corresponding tight frames in $\mathcal{A}(\mathcal{H}_1)$ and $\mathcal{A}(\mathcal{H}_2)$.
Then $\{W_{jk}^1\!\otimes\!W_{\ell m}^2\}$ is a tight frame in $\mathcal{A}(\mathcal{H}_1\!\otimes\!\mathcal{H}_2)$, and
any $A$ from $\mathcal{A}(\mathcal{H}_1\!\otimes\!\mathcal{H}_2)$ admits the representation
\begin{equation}
A=\sum_{j,k=0}^{r_1}\sum_{\ell,m=0}^{r_2} \mathcal{W}_{{}_A}(j,\ell,k,m)W_{jk}^1\!\otimes\!W_{\ell m}^2,
\end{equation}
where
\begin{equation} 
\begin{array}{l}
\mathcal{W}_{{}_A}\!:\!\{0,1,...,r_1\}\!\times\!\{0,1,...,r_2\}\!\times\!\{0,1,...,r_1\}\!\times\!\{0,1,...,r_2\}\!\rightarrow\!\mathbb{R},\\[2mm]
\mathcal{W}_{{}_A}(j,\ell,k,m)\!=\!\mathrm{Tr}(A(W_{jk}^1\!\otimes\!W_{\ell m}^2)),
\end{array}
\end{equation}
is a discrete version of the Wigner function of $A:\mathcal{H}_1\!\otimes\!\mathcal{H}_2\!\rightarrow \mathcal{H}_1\!\otimes\!\mathcal{H}_2$.\\
Since $\sum_{j=0}^{r_1}W_{jj}^1\!=\!\mathbb{I}_{\mathcal{H}_1}$ and $\sum_{\ell=0}^{r_2}W_{\ell \ell}^2\!=\!\mathbb{I}_{\mathcal{H}_2}$, the Wigner function of the partial trace $\mathrm{Tr}_1A:\mathcal{H}_2\!\rightarrow \mathcal{H}_2$ is
the function
\begin{equation} 
\begin{array}{l}
\mathcal{W}_{{}_{\mathrm{Tr}_1A}}\!:\!\{0,1,...,r_2\}\!\times\!\{0,1,...,r_2\}\!\rightarrow\!\mathbb{R},\\[2mm]
\mathcal{W}_{{}_{\mathrm{Tr}_1A}}(\ell,m)\!=\!\sum_{j=0}^{r_1}\mathcal{W}_{{}_A}(j,\ell,j,m),
\end{array}
\end{equation}
and the  Wigner function of the  partial trace $\mathrm{Tr}_2A:\mathcal{H}_1\!\rightarrow \mathcal{H}_1$ is the function
\begin{equation} 
\begin{array}{l}
\mathcal{W}_{{}_{\mathrm{Tr}_2A}}\!:\!\{0,1,...,r_1\}\!\times\!\{0,1,...,r_1\}\!\rightarrow\!\mathbb{R},\\[2mm]
\mathcal{W}_{{}_{\mathrm{Tr}_2A}}(j,k)\!=\!\sum_{\ell=0}^{r_2}\mathcal{W}_{{}_A}(j,\ell,k,\ell).
\end{array}
\end{equation}
\section{Some examples}
\subsection{Triangular frame representation of qubits}
In the two-dimensional Euclidean space $\mathbb{R}^2$, the vectors
\begin{equation} 
\begin{array}{l}
|v_k\rangle\!=\!\sqrt{\frac{2}{3}}\!\left(\!\!
\begin{array}{c}
\cos\frac{2k\pi}{3}\\[1mm]
\sin\frac{2k\pi}{3}
\end{array}\!\!
\right) , \quad \mbox{where}\ \ k\!\in\!\{0,1,2\}, 
\end{array}
\end{equation}
correspond to the vertices of an equilateral triangle,
but in the complex Hilbert space $\mathbb{C}^2$, they
form a tight frame, 
$\sum\limits_{k=0}^2 |v_k\rangle \langle v_k|\!=\!\mathbb{I}_2.$
The operators $V_{jk}\!=\!|v_j\rangle \langle v_k|$
form a tight frame ,
\ \ 
$
\mathbb{I}_{\mathcal{L}(\mathbb{C}^2)}\!=\!\sum\limits_{j,k=0}^2|V_{jk}\rangle\!\rangle\langle \!\langle V_{jk}|$.
The corresponding nine self-adjoint operators
{\footnotesize 
\begin{equation}\label{3qubitframe}
\begin{array}{lll}
W_{00}\!=\!\!\left(\!
\begin{array}{cc}
\frac{2}{3} & 0 \\[2mm]
0 & 0
\end{array}\!\right)\!, & W_{01}\!=\!\!\left(\!\!
\begin{array}{cc}
-\frac{\sqrt{2}}{3} & \frac{1}{\sqrt{6}} \\[2mm]
\frac{1}{\sqrt{6}} & 0
\end{array}\!\!\right)\!, & \!W_{02}\!=\!\!\left(\!\!
\begin{array}{cc}
-\frac{\sqrt{2}}{3} & \frac{-1}{\sqrt{6}} \\[2mm]
\frac{-1}{\sqrt{6}} & 0
\end{array}\!\!\right)\!,\\[5mm]
W_{10}\!=\!\!\left(\!\!
\begin{array}{cc}
0 & \frac{\rm i}{\sqrt{6}} \\[2mm]
\frac{-{\rm i}}{\sqrt{6}} & 0
\end{array}\!\!\right)\!, & W_{11}\!=\!\!\left(\!\!
\begin{array}{cc}
\frac{1}{6} & \frac{-1}{2\sqrt{3}} \\[2mm]
\frac{-1}{2\sqrt{3}} & \frac{1}{2}
\end{array}\!\!\right)\!, & \!W_{12}\!=\!\!\left(\!\!
\begin{array}{cc}
\frac{1}{3\sqrt{2}} & 0 \\[2mm]
0 & -\frac{1}{\sqrt{2}}
\end{array}\!\!\right)\!,\\[5mm]
W_{20}\!=\!\!\left(\!\!
\begin{array}{cc}
0 & \frac{-{\rm i}}{\sqrt{6}} \\[2mm]
\frac{{\rm i}}{\sqrt{6}} & 0
\end{array}\!\!\right)\!, & W_{21}\!=\!\!\left(\!\!
\begin{array}{cc}
0 & \frac{{\rm i}}{\sqrt{6}}  \\[2mm]
\frac{-{\rm i}}{\sqrt{6}}  & 0
\end{array}\!\!\right)\!, & \!W_{22}\!=\!\!\left(\!\!
\begin{array}{cc}
\frac{1}{6} & \frac{1}{2\sqrt{3}} \\[2mm]
\frac{1}{2\sqrt{3}} & \frac{1}{2}
\end{array}\!\!\right)\!,
\end{array}
\end{equation}
}
\noindent form a tight frame in $\mathcal{A}(\mathbb{C}^2)$, \ 
$
\mathbb{I}_{\mathcal{A}(\mathbb{C}^2)}\!=\!\sum\limits_{j,k=0}^2|W_{jk}\rangle\!\rangle\langle \!\langle W_{jk}|$. \ 
If $A\!\in\!\mathcal{A}(\mathbb{C}^2)$ then
\begin{equation}
A\!=\!\sum\limits_{j,k=0}^2\mathcal{W}_A(j,k)\ {W}_{jk},\quad \mathrm{where}\quad
\begin{array}{l}
\mathcal{W}_A\!:\!\{0,1,2\}\!\times\!\{0,1,2\}\!\rightarrow\!\mathbb{R},\\
\mathcal{W}_A(j,k)\!=\! {\rm Tr}(A\,W_{jk}),
\end{array}
\end{equation}
represents a  finite version of the  Wigner function of $A$.
The operators $W_{jk}$ satisfy $W_{21}\!=\!W_{10}\!=\!-W_{20}$ and 
\begin{equation}\label{RWjkRqubit}
\begin{array}{ll}
W_{11}\!=\!R_{-\frac{\pi}{3}}W_{00}R_{-\frac{\pi}{3}}^\dag,\qquad\quad & 
W_{01}\!=\!R_{\frac{\pi}{3}}W_{12}R_{\frac{\pi}{3}}^\dag ,\\
W_{22}\!=\!R_{\frac{\pi}{3}}W_{00}R_{\frac{\pi}{3}}^\dag, & 
W_{02}\!=\!R_{-\frac{\pi}{3}}W_{12}R_{-\frac{\pi}{3}}^\dag, 
\end{array}
\end{equation}
relations similar to (\ref{RPiR}), where  $R_\alpha\!=\!\left(\begin{array}{rr}
\cos \alpha & -\sin \alpha\\
\sin \alpha & \cos \alpha\end{array}\right).$

\subsection{Tetrahedral frame representation for qutrits}
In the three-dimensional Euclidean space $\mathbb{R}^3$, the vectors
  {\footnotesize 
  \begin{equation} 
|v_0\rangle\!=\!\frac{1}{2}\!\left(\!\!
\begin{array}{r}
-1\\[1mm]
1\\[1mm]
1
\end{array}\!\!
\right)\! , \quad |v_1\rangle\!=\!\frac{1}{2}\!\left(\!\!
\begin{array}{r}
1\\[1mm]
-1\\[1mm]
1
\end{array}\!\!
\right) \!, \quad |v_2\rangle\!=\!\frac{1}{2}\!\left(\!\!
\begin{array}{r}
1\\[1mm]
1\\[1mm]
-1
\end{array}\!\!
\right)\!, \quad
|v_3\rangle\!=\!\frac{1}{2}\!\left(\!\!
\begin{array}{r}
-1\\[1mm]
-1\\[1mm]
-1
\end{array}\!\!
\right)
\end{equation}}
\noindent correspond to the vertices of a regular tetrahedron, but
in the complex Hilbert space $\mathbb{C}^3$, they 
form a tight frame.\\ The operators $V_{jk}\!=\!|v_j\rangle \langle v_k|$
 form a  tight frame in $\mathcal{L}(\mathbb{C}^3)$, \ 
$\mathbb{I}_{\mathcal{L}(\mathbb{C}^3)}\!=\!\sum\limits_{j,k=0}^3|V_{jk}\rangle\!\rangle\langle \!\langle V_{jk}|.$
The corresponding operators
{\footnotesize 
\begin{equation} 
\begin{array}{ll}
W_{00}\!=\!\!\frac{1}{4}\left(\!
\begin{array}{rrr}
1 & -1 & -1 \\
-1 & 1 & 1 \\
-1 & 1 & 1 
\end{array}\!\right)\!, & 
\!W_{01}\!=\!\!\frac{1}{2\sqrt{2}}\left(\!
\begin{array}{rrr}
-1 & 1 & 0 \\
1 & -1 & 0 \\
0 & 0 & 1 
\end{array}\!\right)\\[5mm]
W_{02}\!=\!\!\frac{1}{2\sqrt{2}}\left(\!
\begin{array}{rrr}
-1 & 0 & 1 \\
0 & 1 & 0 \\
1 & 0 & -1 
\end{array}\!\right)\!, & 
\!W_{03}\!=\!\!\frac{1}{2\sqrt{2}}\left(\!
\begin{array}{rrr}
1 & 0 & 0 \\
0 & -1 & -1 \\
0 & -1 & -1 
\end{array}\!\right)\\[5mm]
W_{10}\!=\!\!\frac{1}{2\sqrt{2}}\left(\!
\begin{array}{rrr}
0 & 0 & -\mathrm{i} \\
0 & 0 & \mathrm{i} \\
\mathrm{i} & -\mathrm{i} & 0 
\end{array}\!\right)\!, & 
\!W_{11}\!=\!\!\frac{1}{4}\left(\!
\begin{array}{rrr}
1 & -1 & 1 \\
-1 & 1 & -1 \\
1 & -1 & 1 
\end{array}\!\right)\\[5mm]
W_{12}\!=\!\!\frac{1}{2\sqrt{2}}\left(\!
\begin{array}{rrr}
1 & 0 & 0 \\
0 & -1 & 1 \\
0 & 1 & -1 
\end{array}\!\right)\!, & 
\!W_{13}\!=\!\!\frac{1}{2\sqrt{2}}\left(\!
\begin{array}{rrr}
-1 & 0 & -1 \\
0 & 1 & 0 \\
-1 & 0 & -1 
\end{array}\!\right)\\[5mm]
W_{20}\!=\!\!\frac{1}{2\sqrt{2}}\left(\!
\begin{array}{rrr}
0 & -\mathrm{i} & 0 \\
\mathrm{i} & 0 & -\mathrm{i} \\
0 & \mathrm{i} & 0 
\end{array}\!\right)\!, & 
\!W_{21}\!=\!\!\frac{1}{2\sqrt{2}}\left(\!
\begin{array}{rrr}
0 & \mathrm{i} & -\mathrm{i} \\
-\mathrm{i} & 0 & 0 \\
\mathrm{i} & 0 & 0 
\end{array}\!\right)\\[5mm]
W_{22}\!=\!\!\frac{1}{4}\left(\!
\begin{array}{rrr}
1 & 1 & -1 \\
1 & 1 & -1 \\
-1 & -1 & 1 
\end{array}\!\right)\!, & 
\!W_{23}\!=\!\!\frac{1}{2\sqrt{2}}\left(\!
\begin{array}{rrr}
-1 & -1 & 0 \\
-1 & -1 & 0 \\
0 & 0 & 1 
\end{array}\!\right)\\[5mm]
W_{30}\!=\!\!\frac{1}{2\sqrt{2}}\left(\!
\begin{array}{rrr}
0 & \mathrm{i} & \mathrm{i} \\
-\mathrm{i} & 0 & 0 \\
-\mathrm{i} & 0 & 0 
\end{array}\!\right)\!, & 
\!W_{31}\!=\!\!\frac{1}{2\sqrt{2}}\left(\!
\begin{array}{rrr}
0 & -\mathrm{i} & 0 \\
\mathrm{i} & 0 & \mathrm{i} \\
0 & -\mathrm{i} & 0 
\end{array}\!\right)\\[5mm]
W_{32}\!=\!\!\frac{1}{2\sqrt{2}}\left(\!
\begin{array}{rrr}
0 & 0 & -\mathrm{i} \\
0 & 0 & -\mathrm{i} \\
\mathrm{i} & \mathrm{i} & 0 
\end{array}\!\right)\!, & 
\!W_{33}\!=\!\!\frac{1}{4}\left(\!
\begin{array}{rrr}
1 & 1 & 1 \\
1 & 1 & 1 \\
1 & 1 & 1 
\end{array}\!\right)
\end{array}
\end{equation} }
\noindent form a tight frame in $\mathcal{A}(\mathbb{C}^3)$, \ $\mathbb{I}_{\mathcal{A}(\mathbb{C}^3)}\!=\!\sum\limits_{j,k=0}^3|W_{jk}\rangle\!\rangle\langle \!\langle W_{jk}|.$
If $A\!\in\!\mathcal{A}(\mathbb{C}^3)$ then 
\begin{equation}
A\!=\!\sum\limits_{j,k=0}^3\mathcal{W}_A(j,k)\ {W}_{jk},\quad \mathrm{where}\quad
\begin{array}{l}
\mathcal{W}_A\!:\!\{0,1,2,3\}\!\times\!\{0,1,2,3\}\!\rightarrow\!\mathbb{R},\\[2mm]
\mathcal{W}_A(j,k)\!=\! {\rm Tr}(A\,W_{jk}),
\end{array}
\end{equation}
represents a finite version of the  Wigner function of $A$.
The operators $W_{jk}$ satisfy the relations 
\begin{equation}\label{RWjkRqutrit}
\begin{array}{lll}
W_{00}\!=\!\mathcal{R}_4 W_{33}\mathcal{R}_4^\dag, & 
W_{02}\!=\!\mathcal{R}_1W_{01}\mathcal{R}_1^\dag, & 
W_{20}\!=\!\mathcal{R}_3^\dag W_{10}\mathcal{R}_3,\\
W_{11}\!=\!\mathcal{R}_2^\dag W_{33}\mathcal{R}_2, & 
W_{03}\!=\!\mathcal{R}_3W_{01}\mathcal{R}_3^\dag, & 
W_{21}\!=\!\mathcal{R}_1^\dag W_{10}\mathcal{R}_1,\\
W_{22}\!=\!\mathcal{R}_3W_{33}\mathcal{R}_3^\dag, & 
W_{12}\!=\!\mathcal{R}_4W_{01}\mathcal{R}_4^\dag, & 
W_{30}\!=\!\mathcal{R}_3W_{10}\mathcal{R}_3^\dag,\\
 & 
W_{13}\!=\!\mathcal{R}_2W_{01}\mathcal{R}_2^\dag, & 
W_{31}\!=\!\mathcal{R}_2W_{10}\mathcal{R}_2^\dag,\\
 & 
W_{23}\!=\!\mathcal{R}_0W_{01}\mathcal{R}_0^\dag, & 
W_{32}\!=\!\mathcal{R}_0W_{10}\mathcal{R}_0^\dag,\\
\end{array}
\end{equation}
similar to (\ref{RPiR}), where  $\mathcal{R}_i$ are the rotations
{\footnotesize 
\begin{equation}
\begin{array}{rrr}
\mathcal{R}_0\!=\!\!\left(\!\!
\begin{array}{rrr}
-1 & 0 & 0 \\
0 & 1 & 0 \\
0 & 0 & -1 
\end{array}\!\right), & 
\mathcal{R}_1\!=\!\!\left(\!
\begin{array}{rrr}
0 & 1 & 0 \\
0 & 0 & 1 \\
1 & 0 & 0 
\end{array}\!\right), &
\mathcal{R}_2\!=\!\!\left(\!\!
\begin{array}{rrr}
0 & 1 & 0 \\
0 & 0 & -1 \\
-1 & 0 & 0 
\end{array}\!\right),\\[5mm]
\mathcal{R}_3\!=\!\!\left(\!
\begin{array}{rrr}
0 & 0 & -1 \\
-1 & 0 & 0 \\
0 & 1 & 0 
\end{array}\!\right), &
\mathcal{R}_4\!=\!\!\left(\!
\begin{array}{rrr}
0 & 0 & 1 \\
-1 & 0 & 0 \\
0 & -1 & 0 
\end{array}\!\right). & 
\end{array}
\end{equation}
}
\subsection{Icosahedral frame representation of qutrits}
In the 3-dimensional Euclidean space $\mathbb{R}^3$, the points
\begin{equation} 
\begin{array}{lll}
 \pm (1,\tau ,0),\quad &  \pm (-1,\tau ,0),\quad &  \pm (-\tau ,0,1),\\
 \pm (0,-1,\tau ),\ & \pm (\tau ,0,1),\ & \pm (0,1,\tau ),
\end{array}
\end{equation} 
where $\tau\!=\!\frac{1\!+\!\sqrt{5}}{2}$, are the vertices of a regular icosahedron.
In the complex Hilbert space $\mathbb{C}^3$, the vectors 
{\footnotesize 
  \begin{equation}
  \begin{array}{lll}
|v_0\rangle\!=\!\frac{1}{\sqrt{5\!+\!\sqrt{5}}}\left(\!\!\!
\begin{array}{r}
1\\
\tau\\
0
\end{array}\!\!\!
\right) , \  & |v_1\rangle\!=\!\frac{1}{\sqrt{5\!+\!\sqrt{5}}}\left(\!\!\!
\begin{array}{r}
-1\\
\tau\\
0
\end{array}\!\!\!
\right) ,\ & 
 |v_2\rangle\!=\!\frac{1}{\sqrt{5\!+\!\sqrt{5}}}\left(\!\!\!
\begin{array}{r}
- \tau\\
0\\
1
\end{array}\!\!\!
\right), \\[5mm] 
|v_3\rangle\!=\!\frac{1}{\sqrt{5\!+\!\sqrt{5}}}\left(\!\!\!
\begin{array}{r}
0\\
-1\\
\tau
\end{array}\!\!\!
\right) , \ & 
|v_4\rangle\!=\!\frac{1}{\sqrt{5\!+\!\sqrt{5}}}\left(\!\!\!
\begin{array}{r}
\tau\\
0\\
1
\end{array}\!\!\!
\right) , \ & |v_5\rangle\!=\!\frac{1}{\sqrt{5\!+\!\sqrt{5}}}\left(\!\!\!
\begin{array}{r}
0\\
1\\
\tau
\end{array}\!\!\!
\right)
\end{array}
\end{equation}}
\noindent form a tight frame \cite{Cotfas10}.
The operators $V_{jk}\!=\!|v_j\rangle \langle v_k|$ form a tight frame in 
$\mathcal{L}(\mathbb{C}^3)$, and the corresponding operators $W_{jk}$ form a tight frame in $\mathcal{A}(\mathbb{C}^3)$.\\
\begin{figure*}[t]
\includegraphics[scale=0.9]{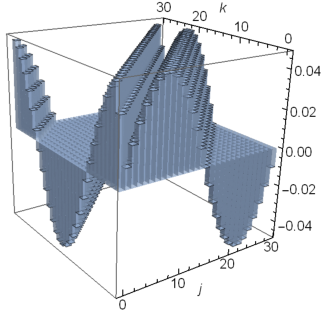}\qquad \quad
\includegraphics[scale=1]{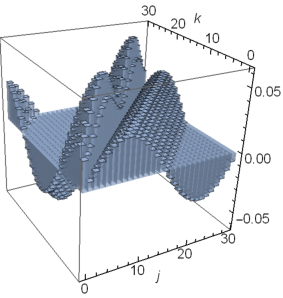}\qquad 
\includegraphics[scale=0.9]{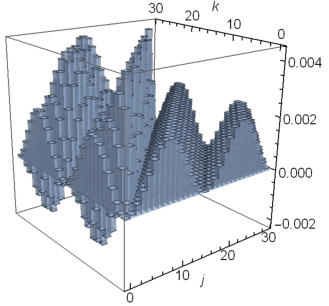}\\
\includegraphics[scale=0.9]{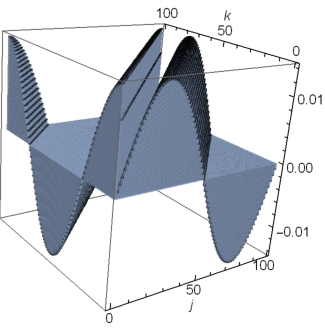}\quad 
\includegraphics[scale=0.9]{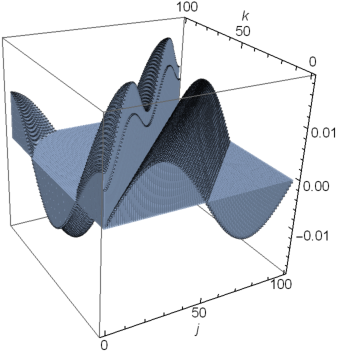}\quad 
\includegraphics[scale=0.9]{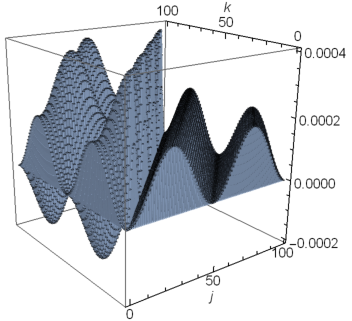}
\caption{Wigner function of three quantum states described by using the  tight frame (\ref{mframe}) with $m\!=\!30$ and $m\!=\!100$:\\   {\it Left:} \ Pure state $\frac{1}{\sqrt{2}}\left(\begin{array}{c}1\\ -\mathrm{i}\end{array}\right)$; \quad {\it Center:} \ Mixed state $\frac{1}{3}\left(\begin{array}{rr}1 & \mathrm{i}\\ -\mathrm{i} & 2\end{array}\right)$; \quad {\it Right:} \ Bell state $\frac{1}{\sqrt{2}}\left(\left(\!\begin{array}{c}1\\ 0\end{array}\!\right)\!\otimes\!\left(\!\begin{array}{c}0\\ 1\end{array}\!\right)+\left(\!\begin{array}{c}0\\ 1\end{array}\!\right)\!\otimes\!\left(\!\begin{array}{c}1\\ 0\end{array}\!\right)\right)$.}
\end{figure*}
\section{Some applications}

\subsection{Tomography of a quantum state}
If there exists a rotation $R$ such that $W_{j'k'}\!=\!R\, W_{jk}\, R^\dag$, then
\begin{equation} 
\mathcal{W}_\varrho(j',k')\!=\!\mathrm{Tr}(\varrho \, R\, W_{jk}\, R^\dag)\!=\!\mathrm{Tr}(R^\dag\, \varrho \, R\, W_{jk}),
\end{equation} 
that is $\mathcal{W}_\varrho(j',k')$ can be measured either by rotating
the quantum system or the apparatus used for $\mathcal{W}_\varrho(j,k)$.
The relations (\ref{RWjkRqubit}) and (\ref{RWjkRqutrit}) show that three experimental setups are
sufficient for the measurement of the Wigner function of a state in the considered particular cases concerning the qubit and qutrit.

We are mainly interested in the cases when the starting frame $\{|v_j\rangle\}$
is generated by the action of a finite group $G$ of rotations. In the case of such a $G$-frame, for any $j$ there exists a rotation $R_j\!\in\!G$ such that $|v_j\rangle\!=\!R_j|v_0\rangle$. \ We have \ 
$W_{jj}\!=\!R_j\, W_{00}\, R_j^\dag$ for any $j$, and if the rotation $R_m\!\in\!G$ is such that $R_m\!=\!R_j^\dag R_k$, then
\begin{equation} \label{rotqudit}
\begin{array}{l}
W_{jk}\!=\!R_jW_{0m}R_j^\dag\quad \mbox{for}\quad  j\!<\!k,\\
W_{jk}\!=\!R_jW_{m0}R_j^\dag\quad \mbox{for}\quad  j\!>\!k
\end{array}
\end{equation} 
So, certain $W_{jk}$ are related through rotations.

The presented frame representation can be regarded as a kind of discrete version of the continuous Wigner function of qudits described in \cite{Tilma16,Rundle17,Koczor20,Rundle21}.
In the case of a continuous phase space, the Wigner function is reconstructed by using an expansion in spherical harmonics  from a finite number of measurements. 
If we use the frame representation presented above, then for tomography of a quantum state $\varrho $, we have to measure all the values of the Wigner function (mean values of all $W_{jk}$ in the state $\varrho $). The relations (\ref{rotqudit})  show that the number of necessary experimental setups remains relatively small when we increase the number of the elements of the frame.

\subsection{Visualization of quantum states}
In the case of the qubit, the vectors of a starting frame of the form
\begin{equation}\label{mframe} 
\begin{array}{l}
|v_k\rangle\!=\!\sqrt{\frac{2}{m}}\!\left(\!\!
\begin{array}{c}
\cos\frac{2k\pi}{m}\\[1mm]
\sin\frac{2k\pi}{m}
\end{array}\!\!
\right) , 
\end{array}
\end{equation}
where $k\!\in\!\{0,1,2,...,m\!-\!1\}$, can be arranged in a natural way in the sequence
$|v_0\rangle$, $|v_1\rangle$,...,$|v_{m-1}\rangle$, and the set
$\{0,1,...,m\!-\!1\}\!\times\!\{0,1,...,m\!-\!1\}$ regarded as a natural phase space. This allows an intuitive representation of a state.
Since the tensor product of two tight frames is a tight frame, some significant
intuitive representations can be obtained for a composite system of $N$ qubits by using the equal-coordinate slice (similar to equal-angle slice used in \cite{Tilma16,Rundle17,Koczor20,Rundle21}). In figure 1, we represent the Wigner functions $\mathcal{W}_m$ corresponding to three quantum states in the representations defined by the regular tight frame (\ref{mframe}) with $m\!=\!30$ and $m\!=\!100$.\\
\\
If the dimension of the Hilbert space $\mathcal{H}$ is greater than 2, then the vectors $|v_k\rangle$ are distributed on a sphere in $\mathcal{H}$, and the order in the sequence $|v_0\rangle$, $|v_1\rangle$,...,$|v_{r}\rangle$ is chosen arbitrary. So, we can not use our description $\{0,1,...,r\}\!\times\!\{0,1,...,r\}$ of the phase space for direct intuitive representations.
 \begin{table}
\caption{\label{purity}
Dependence on $m$ of $\mathcal{N}_m$ and $\mathcal{C}_m$  in two particular cases.
}
\begin{indented}
\item[]\begin{tabular}{@{}cccccccc}
\br
Qubit state & $m$ & $\mathcal{N}_m$ & $\mathcal{C}_m$ & Qubit state & $m$ & $\mathcal{N}_m$ & $\mathcal{C}_m$  \\[2mm]
\mr
$\left(\!\!\!\begin{array}{r}\frac{1}{\sqrt{2}}\\[2mm] -\frac{\mathrm{i}}{\sqrt{2}}\end{array}\!\!\!\right)$  & $\begin{array}{c}
3\\
4\\
5\\
6\\
7\\
8\\
9\\
10\\
15\\
20\\
30\\
40\\
50\\
60\\
70\\
80\\
90\\
100
\end{array}$   & $\begin{array}{c}
0.3035\\
0.1875\\
0.2595\\
0.2388\\
0.2490\\
0.2263\\
0.2449\\
0.2387\\
0.2409\\
0.2367\\
0.2387\\
0.2382\\
0.2387\\
0.2385\\
0.2387\\
0.2386\\
0.2387\\
0.2386
\end{array}$      & $\begin{array}{c}
0.6439\\
0.5303\\
0.7514\\
0.7618\\
0.7955\\
0.7651\\
0.8194\\
0.8221\\
0.8523\\
0.8575\\
0.8759\\
0.8807\\
0.8858\\
0.8877\\
0.8900\\
0.8910\\
0.8923\\
0.8929
\end{array}$     & 
$\left(\!\!\!\begin{array}{rr}\frac{1}{3} & \frac{\mathrm{i}}{3}\\[2mm]
-\frac{\mathrm{i}}{3} & \frac{2}{3}\end{array}\!\!\!\right)$ & $\begin{array}{c}
3\\
4\\
5\\
6\\
7\\
8\\
9\\
10\\
15\\
20\\
30\\
40\\
50\\
60\\
70\\
80\\
90\\
100
\end{array}$    & $\begin{array}{c}
0.2769\\
0.1250\\
0.1921\\
0.1840\\
0.1770\\
0.1587\\
0.1713\\
0.1686\\
0.1654\\
0.1618\\
0.1631\\
0.1623\\
0.1627\\
0.1624\\
0.1626\\
0.1624\\
0.1625\\
0.1624
\end{array}$     & $\begin{array}{c}
0.5078\\
0.4124\\
0.6091\\
0.6257\\
0.6569\\
0.6523\\
0.6822\\
0.6798\\
0.7144\\
0.7239\\
0.7380\\
0.7443\\
0.7483\\
0.7507\\
0.7525\\
0.7538\\
0.7548\\
0.7556
\end{array}$  \\
\br
\end{tabular}
\end{indented}
\end{table}

\subsection{A frame version of volume of the negative part of Wigner function}
The frame version of the volume of the negative part of Wigner function \cite{Kenfack04} can be defined as
 {\scriptsize
\begin{equation}\label{m-negativity} 
\mathcal{N}_m\!=\!\frac{\sum\limits_{j,k}(|\mathcal{W}_m(j,k)|-\mathcal{W}_m(j,k))}{2\, m^2\, \max\limits_{j,k}|\mathcal{W}_m(j,k)|}.
\end{equation}
}
By taking into consideration the behaviour of frame representation for $m\rightarrow \infty$ (see Fig. 1 and Table 1), we can define the volume  of the negative part of Wigner function in a new way, namely as
\begin{equation}\label{m-negativity} 
\mathcal{N}\!=\!\lim\limits_{m\rightarrow \infty} \mathcal{N}_m.
\end{equation}
This new parameter  does not depend on the particular frame representation (\ref{mframe}) we choose.
Two examples are presented in Table 1.

\subsection{Coherence of a state with respect to a frame}

In the case of a frame representation, for any quantum state $\varrho$, we have
\begin{equation}
\sum_{j=0}^r  |v_j\rangle \langle v_j|\!=\!\mathbb{I}_{\mathcal{H}}\quad \Rightarrow \quad 
\sum_{j=0}^r  W_{jj}\!=\!\mathbb{I}_{\mathcal{H}}\quad \Rightarrow \quad \sum_{j=0}^r  \mathcal{W}_\varrho(j,j)\!=\!1.
\end{equation}
The parameter
\begin{equation}
\mathcal{C}_m\!=\!\frac{1}{m}\sum_{j\neq k}  |\mathcal{W}_\varrho(j,k)|
\end{equation}
may contain some information concerning the coherence of a qubit quantum state $\varrho$.
The {\em frame coherence of a qubit} might be defined as the limit
\begin{equation}
\mathcal{C}\!=\!\lim\limits_{m\rightarrow \infty} \mathcal{C}_m,
\end{equation}
which seems to exist for any  state $\varrho$ (see Table 1), and  does not depend on the particular frame representation (\ref{mframe}) we choose.
\subsection{Qubit regarded as an orthogonal projection of qutrit}

The use of the triangular frame representation of qubit makes more transparent the relation qubit-qutrit.
The tight frame 
\begin{equation}
|u_0\rangle\!=\!\left( 
\begin{array}{c}
\sqrt{\frac{2}{3}}\\[2mm]
0
\end{array}\right),\quad |u_1\rangle\!=\!\left( 
\begin{array}{r}
-\frac{1}{\sqrt{6}}\\[1mm]
\frac{1}{\sqrt{2}}
\end{array}\right),\quad |u_2\rangle\!=\!\left( 
\begin{array}{r}
-\frac{1}{\sqrt{6}}\\[1mm]
-\frac{1}{\sqrt{2}}
\end{array}\right),
\end{equation}
allows us to identify the Hilbert space $\mathbb{C}^2$ of qubit with the subspace
\begin{equation}
\mathcal{S}\!=\!\{\ \alpha_0|w_0\rangle\!+\!\alpha_1|w_1\rangle\ \ |\ \ \alpha_0,\, \alpha_1\!\in\!\mathbb{C}\ \}
\end{equation}
of the Hilbert space $\mathbb{C}^3$ of qutrit by using the linear isometry (see (\ref{embedding}))
\begin{equation}
\mathbb{C}^2\longrightarrow\mathcal{S}:\quad 
\alpha_0|e_0\rangle\!+\!\alpha_1|e_1\rangle\mapsto 
\alpha_0|w_0\rangle\!+\!\alpha_1|w_1\rangle,
\end{equation}
where
\begin{equation}\fl
|e_0\rangle\!=\!\left( 
\begin{array}{c}
1\\[2mm]
0
\end{array}\right),\quad 
|e_1\rangle\!=\!\left( 
\begin{array}{c}
0\\[2mm]
1
\end{array}\right),\quad 
|w_0\rangle\!=\!\left(\!\! 
\begin{array}{r}
\sqrt{\frac{2}{3}}\\[2mm]
-\frac{1}{\sqrt{6}}\\[2mm]
-\frac{1}{\sqrt{6}}
\end{array}\!\!\right),\quad 
|w_1\rangle\!=\!\left( \!\!
\begin{array}{r}
0\\[1mm]
\frac{1}{\sqrt{2}}\\[2mm]
-\frac{1}{\sqrt{2}}
\end{array}\!\!\right).
\end{equation}
Since $\{ |w_0\rangle,\,|w_1\rangle \}$ is an orthonormal basis of $\mathcal{S}$, the orthogonal projector corresponding to $\mathcal{S}$ is $P:\mathbb{C}^3\longrightarrow\mathcal{S}$,
\begin{equation}
P\!=\!|w_0\rangle\langle w_0|\!+\!|w_1\rangle\langle w_1|\!=\!\left(\!\!\begin{array}{rrr}
\frac{2}{3} & -\frac{1}{3} & -\frac{1}{3}\\[2mm]
-\frac{1}{3} & \frac{2}{3} & -\frac{1}{3}\\[2mm]
-\frac{1}{3} & -\frac{1}{3} & \frac{2}{3}
\end{array}\!\!\right).
\end{equation}
The orthogonal projector
\begin{equation}
\mathcal{A}(\mathbb{C}^3)\longrightarrow \mathcal{A}(\mathcal{S})\!\equiv\!\mathcal{A}(\mathbb{C}^2):\quad A\mapsto PAP
\end{equation}
is a positive map
\[
A\geq 0\quad \Rightarrow \quad \langle x,PAPx\rangle \!=\!\langle Px,APx\rangle \!\geq\!0.
\]
For any $A\!\in\!\mathcal{A}(\mathbb{C}^3)$, the matrix of 
$PAP$ in the basis $\{|e_0\rangle, \ |e_1\rangle\}$ is 
$A'\!=\!LPAPL^T$, where
\begin{equation}
L\!=\!\left( 
\begin{array}{rrr}
\sqrt{\frac{2}{3}} & 
-\frac{1}{\sqrt{6}} &
-\frac{1}{\sqrt{6}}\\[2mm]
0 &
\frac{1}{\sqrt{2}} &
-\frac{1}{\sqrt{2}}
\end{array}\right) 
\end{equation}
satisfies $LL^T\!=\!\mathbb{I}_{\mathbb{C}^2}$ and $L^TL\!=\!P$.
The tight frame (\ref{3qubitframe}) of $\mathcal{A}(\mathbb{C}^2)$ is the orthogonal projection of the orthonormal basis
{\footnotesize 
\begin{equation}
\begin{array}{ccc}
E_{00}\!=\!\!\left(\!
\begin{array}{rrr}
1 & 0 & 0\\[1mm]
0 & 0 & 0 \\[1mm]
0 & 0 & 0 
\end{array}\!\right)\!, & E_{01}\!=\!\!\left(\!
\begin{array}{rrr}
0 & \frac{1}{\sqrt{2}} & 0\\[1mm]
\frac{1}{\sqrt{2}} & 0 & 0 \\[1mm]
0 & 0 & 0 
\end{array}\!\right)\!, & \!E_{02}\!=\!\!\left(\!
\begin{array}{rrr}
0 & 0 & \frac{1}{\sqrt{2}}\\[1mm]
0 & 0 & 0 \\[1mm]
\frac{1}{\sqrt{2}} & 0 & 0 
\end{array}\!\right)\!,\\[6mm]
E_{10}\!=\!\!\left(\!
\begin{array}{rrr}
0 & \frac{\mathrm{i}}{\sqrt{2}} & 0\\[1mm]
-\frac{\mathrm{i}}{\sqrt{2}} & 0 & 0 \\[1mm]
0 & 0 & 0 
\end{array}\!\right)\!, & E_{11}\!=\!\!\left(\!
\begin{array}{rrr}
0 & 0 & 0\\[1mm]
0 & 1 & 0 \\[1mm]
0 & 0 & 0 
\end{array}\!\right)\!, & \!E_{12}\!=\!\!\left(\!
\begin{array}{rrr}
0 & 0 & 0\\[1mm]
0 & 0 & \frac{1}{\sqrt{2}} \\[1mm]
0 & \frac{1}{\sqrt{2}} & 0 
\end{array}\!\right)\!,\\[6mm]
E_{20}\!=\!\!\left(\!
\begin{array}{rrr}
0 & 0 & \frac{\mathrm{i}}{\sqrt{2}}\\[1mm]
0 & 0 & 0 \\[1mm]
-\frac{\mathrm{i}}{\sqrt{2}} & 0 & 0 
\end{array}\!\right)\!, & E_{21}\!=\!\!\left(\!
\begin{array}{rrr}
0 & 0 & 0\\[1mm]
0 & 0 & \frac{\mathrm{i}}{\sqrt{2}} \\[1mm]
0 & -\frac{\mathrm{i}}{\sqrt{2}} & 0 
\end{array}\!\right)\!, & \!E_{22}\!=\!\!\left(\!
\begin{array}{rrr}
0 & 0 & 0\\[1mm]
0 & 0 & 0 \\[1mm]
0 & 0 & 1 
\end{array}\!\right)\!,
\end{array}
\end{equation}
}
\noindent of $\mathcal{A}(\mathbb{C}^3)$, that is 
\begin{equation}
W_{jk}\!=\!LPE_{jk}PL^T,\qquad \mathrm{for\ any} \qquad 
i,j\!\in\!\{0,1,2\}.
\end{equation}

The map $\mathcal{A}(\mathbb{C}^3)\!\rightarrow \mathcal{A}(\mathbb{C}^2):A\mapsto A'\!=\!LPAPL^T$ from the qutrit to qubit is surjective but not injective. Since 
$W_{jk}\!=\!LPE_{jk}PL^T\Rightarrow L^TW_{jk}L\!=\!PE_{jk}P$, we get
\[
\begin{array}{r}
\mathcal{W}_{A'}(j,k)\!=\!\mathrm{Tr}(W_{jk}A')\!=\!
\mathrm{Tr}(W_{jk}LPAPL^T)\!=\!
\mathrm{Tr}(L^TW_{jk}LPAP)\\
=\!\mathrm{Tr}(PE_{jk}PPAP)\!=\!
\mathrm{Tr}(E_{jk}PAP),
\end{array}
\]
that is , the Wigner function corresponding to $A'$ is
\begin{equation}
\mathcal{W}_{A'}(j,k)\!=\!\mathrm{Tr}(E_{jk}PAP).
\end{equation}

\subsection{Gaussian states of qubit}
\begin{figure*}[t]
\includegraphics[scale=1.1]{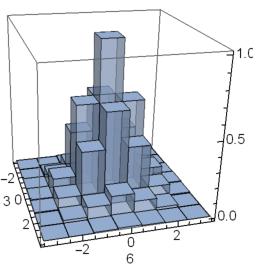} \quad
\includegraphics[scale=1.1]{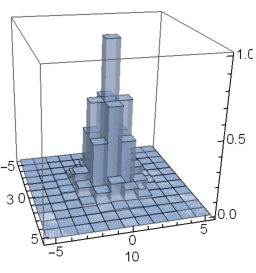}\quad 
\includegraphics[scale=1.1]{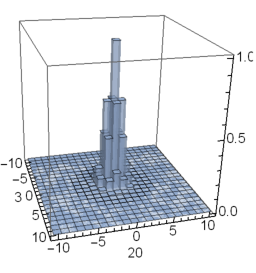}
\caption{The discrete Gaussian function $f_{1/2} (j,k)\!=\!\mathrm{e}^{-(j^2+k^2)/2}$ in the cases $m\!=\!7$,
$m\!=\!11$ and $m\!=\!21$.}
\end{figure*}
In order to simplify the notations, in this section, we consider only tight frames $\{W_{jk}\}$ obtained by starting from frames (\ref{mframe}) with odd $m\!=\!2 \ell \!+\!1$.
Instead of the phase space $\{0,1,...,m\!-\!1\}\!\times\{0,1,...,m\!-\!1\}$ we use $\{-\ell ,\ell\!+\!1,...,\ell\!-\!1,\ell\}\!\times\{-\ell ,\ell\!+\!1,...,\ell\!-\!1,\ell\}$, and define
\begin{equation}
W_{jk}\!=\!W_{j(\mathrm{mod}\, m)\, k(\mathrm{mod}\, m)}
\end{equation}
for any $j,k\!\in\!\{-\ell ,\ell\!+\!1,...,\ell\!-\!1,\ell\}.$ A function of the form (see Fig.2)
\begin{equation}
f_\kappa:\{-\ell ,\ell\!+\!1,...,\ell\!-\!1,\ell\}\!\times\{-\ell ,\ell\!+\!1,...,\ell\!-\!1,\ell\}\!\rightarrow \!\mathbb{R}, \qquad 
f_\kappa (j,k)\!=\!\mathrm{e}^{-\kappa (j^2+k^2)},
\end{equation}
with $\kappa \!\in\!(0,\infty)$, is a discrete version of the Gaussian function $g_\kappa\!:\!\mathbb{R}\!\times\!\mathbb{R}\!\rightarrow\!\mathbb{R},\ \ g_\kappa (x,y)\!=\!\mathrm{e}^{-\kappa (x^2+y^2)}$.\\
A quantum state of the form
\begin{equation}
\varrho_\kappa\!:\!\mathbb{C}^2\!\rightarrow\!\mathbb{C}^2,\qquad \varrho_\kappa=\frac{\sum\limits_{j,k=-\ell}^\ell \mathrm{e}^{-\kappa (j^2+k^2)}\, W_{jk}}{\mathrm{Tr}(\sum\limits_{j,k=-\ell}^\ell \mathrm{e}^{-\kappa (j^2+k^2)}\, W_{jk})}
\end{equation}
can be regarded as a {\em Gaussian state of the qubit}. 
We do not know which restrictions $\ell$ and $\kappa $ have to satisfy in order to have $\varrho_\kappa\!\geq\!0$.
In the investigated cases (see Table 2), the operators $\varrho_\kappa$ defined in this way are density operators.
We think that the Gaussian states of the qubit may have some interesting properties. Our intention is to show that the frame representation offers the possibility to define  new parameters describing the quantum states as well as some "special" states.
 \begin{table}
\caption{\label{eigval}
Eigenvalues of $\varrho_\kappa$ in certain particular cases.
}
\begin{indented}
\item[]\begin{tabular}{@{}cccccc}
\br
\ \ $\kappa$ & \ \ m &  Eigenvalues of $\varrho_\kappa$ & $\kappa$ & \ \ m &  Eigenvalues of $\varrho_\kappa$\\[2mm]
\mr
$\frac{1}{2}$ & $\begin{array}{r}
3\\
5\\
7\\
9\\
11\\
21
\end{array}$   & 
 $\begin{array}{l}
0.935639\\
0.865239\\
0.949816\\
0.977093\\
0.987222\\
0.997449
\end{array}$\quad      
$\begin{array}{l}
0.0643606\\
0.134761\\
0.0501843\\
0.0229066\\
0.0127781\\
0.00250995
\end{array}$ & 
$1$ & $\begin{array}{r}
3\\
5\\
7\\
9\\
11\\
21
\end{array}$   & 
 $\begin{array}{l}
0.928317\\
0.949233\\
0.976296\\
0.986562\\
0.991348\\
0.997774
\end{array}$\quad      
$\begin{array}{l}
0.0716833\\
0.0507668\\
0.023707\\
0.0134375\\
0.0086524\\
0.00222598
\end{array}$\\
\br
\end{tabular}
\end{indented}
\end{table}

\subsection{Error correction of measurements}
The redundant information introduced by passing from orthogonal bases to tight frames may allow us to eliminate certain errors, to increase the precision of measurements \cite{DeBrota20}. For example, in the odd dimensional case, if $\varrho$ is a quantum state, $W_{jk}$ are the operators describing the qudit in a frame representation and $\Pi(j,k)$ are the operators (\ref{Pijk}), then it is known \cite{Vourdas17} that the error 
\begin{equation}
\parallel \varrho - \sum_{j,k=0}^r (\mathrm{Tr}(\varrho\, W_{jk})\!+\!\lambda_{jk})\,W_{jk}\parallel
\end{equation}
is generally smaller than the error
\begin{equation}
\parallel \varrho - \sum_{j,k=-s}^s (\frac{1}{d}\mathrm{Tr}(\varrho\, \Pi(j,k))\!+\!\mu_{jk})\,\Pi(j,k)\parallel
\end{equation}
for any random numbers $\lambda_{jk}$ and $\mu_{jk}$ lying in a small neighborhood $(-\varepsilon,\varepsilon)$ of $0$. In order to increase the precision of a measurement, we have to start from a frame containing more vectors.

\subsection{Error detection and correction}
Only a part of the operators $W_{jk}$ used in a frame representation are linearly independent.
For example, in the case of the triangular frame representation of qubits, any five of the eight operators  $W_{jk}$ 
from (\ref{3qubitframe}) are linearly dependent. This means that, for any five of the eight values 
$\mathcal{W}_\varrho(j,k)$, a certain linear combination must be null.
If only a small part of $\mathcal{W}_\varrho(j,k)$ contain errors, these errors can be detected and corrected
by using the $\frac{8!}{3!\, 5!}\!=\!56$ restrictions  the values of the Wigner function of any 
state $\varrho$ must satisfy.

\section{Concluding remarks}
The method to obtain a tight frame $\{W_{jk}\}$ of $\mathcal{A}(\mathcal{H})$ by starting from a tight frame $\{|v_j\rangle\}$ of $\mathcal{H}$ seems (to our knowledge) to be new. Based on it, we have obtained some explicit representations for qubits and qutrits. In the case of coherent states, the representation of a pure state as a linear combination of coherent states is not unique, but there exists a standard representation based on the resolution of the identity very useful in applications.
In a very similar way, the representation of a quantum state or observable as a linear combination of $W_{jk}$  is not unique, but there exists a representation based on the resolution of the identity. By using this standard representation, we define a more general version of the discrete Wigner function. We think that the presented approach may be useful in the investigation of the properties of the quantum systems involving qubits and qutrits.

\section*{References}

\end{document}